\documentstyle[prb,tighten,eqsecnum,aps]{revtex}
\begin{document}
\title{$\eta$-Exponents in the one dimensional antiferromagnetic
Heisenberg model with next to nearest neighbour coupling}
\author{C. Gerhardt, A. Fledderjohann, E. Aysal,
        K.-H. M\"utter\footnote{e-mail:muetter@wpts0.physik.uni-wuppertal.de} 
        }
\address{Physics Department, University of Wuppertal, 42097 Wuppertal, Germany}
\author{J.F. Audet, H. Kr\"oger}
\address{Department of Physics, Universit\'e Laval,
Quebec, Canada}

\date{\today}
\maketitle
%\twocolumn
%%%%%%%%%%%%%%%%%%%%%%%%%%%%%%%%%%%%%%%%%%%%%%%%%%%%%%%%%%%%%%%%%%%%%%%%%%%%%%%%
%
%
\begin{abstract}
We investigate the critical exponents $\eta_3(\alpha,M)$, $\eta_1(\alpha,M)$
associated with the singularities in the longitudinal and transverse
structure factors of the one dimensional antiferromagnetic Heisenberg
model with nearest $(J_1)$ and next to nearest $(J_2)$
neighbour coupling of relative strength $\alpha=\frac{J_2}{J_1}$ and an
external field $B$ with magnetization $M(B)$. 
\end{abstract}
%%%%%%%%%%%%%%%%%%%%%%%%%%%%%%%%%%%%%%%%%%%%%%%%%%%%%%%%%%%%%%%%%%%%%%%%%%%%%%%%

\section{Introduction}
%There are various possibilities to destroy antiferromagnetic order in
%quantum spin systems. Next to nearest neighbour interactions have this
%property as well as the presence of a uniform external field $B$. \\
In this paper we continue the investigation of the one dimensional
spin $\frac{1}{2}$ Heisenberg model
\begin{equation}
\label{a1}
  H = 2 \sum_{x=1}^N (\vec{S}(x) \vec{S}(x+1) + \alpha \vec{S}(x)
\vec{S}(x+2)) + 2 B \sum_{x=1}^N S_3 (x)
\end{equation}
with next to nearest neighbour coupling parameter $\alpha$ and external field $B$. 

Let us briefly summarize those results [\onlinecite{1}] -
[\onlinecite{8}] relevant for our later
investigation. \\
[5mm]
In the absence of an external field ($B=0$), the ground state of the
model is a singlet ($S=0$) state with momentum $p_0$, where $p_0=0$
for $N=4,8,12, ...$ and $p_0=\pi$ for $N=6,10,14,...$. This statement
holds at least in the interval $-\frac{1}{2} < \alpha <
\frac{1}{2}$. At $\alpha = \frac{1}{2}$ the model reduces to the
Majumdar Ghosh model \cite{9} with degenerate dimer ground states. \\
The quantum numbers of the first excited state change with $\alpha$:
There is a triplet $S=1$ state for $ 0 < \alpha < \alpha_c =
0.241...$ and a singlet $(S=0)$ state for $\alpha_c < \alpha <
\frac{1}{2}$. The momentum of the first excited state is
$p_1=p_0+\pi$. Moreover there is no gap in the 'spinfluid' phase and a
gap in the 'dimer' phase $\alpha > \alpha_c$. 
The structure of these two phases can be exploited by means of the
static and dynamical correlation functions of appropriate
operators. In the spinfluid phase $0 < \alpha < \alpha_c$ the $\Delta S=1$
operator
\begin{equation}
S_3(p) = \frac{1}{\sqrt{N}}\sum_x e^{ipx} S_3(x)
\label{a2}
\end{equation} 
generates the transition from the singlet to the triplet excited
state. To examine the dimer phase $\alpha > \alpha_c$ we need a
$\Delta S=0$ operator. These transitions are generated by the dimer
operator 
\begin{equation}
D(p) = \frac{1}{\sqrt{N}} 
\sum_x e^{ipx} (\vec{S}(x)\vec{S}(x+1) -<\vec{S}(x)\vec{S}(x+1)>)
\label{a3}
\end{equation}
The corresponding static structure factors
\begin{equation}
S_3(\alpha,p,N) = <S^+_3(p) S_3(p)> \mbox{ and } D(\alpha,p,N)=<D^+(p) D(p)>
\label{a4}
\end{equation} 
behave as follows for  $N \to \infty$: \\
$S_3(\alpha,p=\pi,N)$  diverges logarithmically for $\alpha \le \alpha_c$
but stays finite for $\alpha > \alpha_c$. \\
$D(\alpha,p=\pi,N)$  diverges with a power depending on $\alpha$ for
$\alpha >  \alpha_c$ and stays finite for  $\alpha < \alpha_c$. 
The power behaviour degenerates to a logarithmic behaviour for
$\alpha = \alpha_c$. \\
[5mm]
In the presence of an external field $B$ the ground state of the
model has total spin $S=M \cdot N$, where $M=M(\alpha,B)$ is the
magnetization. The behaviour of the magnetization curve $M(\alpha,B)$
near saturation $B \to B_s$, $M \to \frac{1}{2}$ changes with
$\alpha$ \hspace{1mm} \cite{10}.\\
 At $\alpha=0$ it is known to develop a square root
singularity \cite{11}:
\begin{equation}
M(\alpha=0, B) \to \frac{1}{2} - \frac{1}{\pi} (B_s-B)^{\frac{1}{2}},
\mbox{ for } B \to B_s ,
\label{a5}
\end{equation}
whereas the numerical data for $\alpha=\frac{1}{4} (N \le 28)$ support
a quartic root singularity for $\alpha=\frac{1}{4}$ \hspace{1mm} \cite{10}:
\begin{equation}
M(\alpha=\frac{1}{4}, B) \to \frac{1}{2} - \frac{1}{2 \epsilon_4^{\frac{1}{4}}}
(B_s-B)^{\frac{1}{4}} \hspace{5mm} \epsilon_4=1.70(5) ,
\mbox{ for } B \to B_s .
\label{a6}
\end{equation}
The gap in the dimer phase $\alpha > \alpha_c$ appears in the low
field behaviour of the magnetization curve: 
\begin{equation}
M(\alpha, B) \to 0 , \hspace{5mm} B<B_c(\alpha) , \hspace{5mm} \alpha
> \alpha_c .
\label{a7}
\end{equation} 
Note that the model with $B>B_c(\alpha)$ is gapless, provided that
there are no 'plateaus' $(M(\alpha,B) = const, B_{1c} \le B \le
B_{2c})$ in the magnetization curve. \\
The momentum of the ground state
$p_S$ follows Marshall's sign rule \cite{12}:
\begin{equation}
p_S = 0 \hspace{2mm} \mbox{ for } \hspace{2mm} 2S+N = 4 n,  \hspace{5mm}
p_S = \pi \hspace{2mm} \mbox{ for } \hspace{2mm}
2S+N = 4 n + 2 \hspace{3mm} 
\label{a8}
\end{equation}
for $ 0 < \alpha < \frac{1}{4}$. \\
The singularities in the static structure factors change, if we switch
on an external field: The transverse structure factor at $p=\pi,
\alpha=0$
\begin{equation}
S_1(\alpha,p=\pi,M,N) \approx B_1(\alpha,M) N^{1-\eta_1(\alpha,M)} + A_1(\alpha,M)
\label{a9}
\end{equation} 
diverges with a field dependent critical exponent
$\eta_1(\alpha,M)$. $\eta_1(\alpha=0,M)$ has been calculated in
Ref.\onlinecite{13} by means of the
Bethe Ansatz. A second, weaker singularity, which moves with the
external field appears at the softmode momentum $p=p_1(M)=2 \pi M$. To
our knowledge, the positions of both singularities do not depend on
$\alpha$. The critical exponent $\eta_1(M,\alpha)$ however does. It
has been found to be \cite{14}
\begin{equation}
\eta_1(\alpha=0,M=\frac{1}{4}) = 0.65 \hspace{2mm} \mbox{and} \hspace{2mm}
\eta_1(\alpha=\frac{1}{4},M=\frac{1}{4}) = 1.16
\label{a10}
\end{equation}
The longitudinal structure factor at $p=\pi$ ($M>0$) stays finite for $\alpha
< \frac{1}{4}$ but develops a singularity at the field dependent
softmode momentum $p=p_3(M)=\pi(1-2M)$:
\begin{equation}
S_3(\alpha,p=p_3(M),M,N) \approx B_3(\alpha,M) N^{1-\eta_3(\alpha,M)}
+  A_3(\alpha,M)
\label{a11}
\end{equation} 
Again, the position of the singularity does not depend on $\alpha$,
whereas the critical exponent $\eta_3(\alpha,M)$ changes drastically
with $\alpha$:
\begin{equation}
\eta_3(\alpha=0,M=\frac{1}{4}) = 1.50 \hspace{2mm} \mbox{and} \hspace{2mm}
\eta_3(\alpha=\frac{1}{4},M=\frac{1}{4}) = 0.84
\label{a12}
\end{equation}
The soft mode singularity at $p=p_3(M)$ is also found in the dimer
structure factor $D(\alpha,p,N)$ defined in (\ref{a3}) and
(\ref{a4}). A finite size analysis of the type (\ref{a11}) for
$D(\alpha,p,N)$ yields for the critical exponents \cite{14}:
\begin{equation}
 \eta_D(\alpha=0, M=\frac{1}{4}) = 1.49 \hspace{5mm}
 \eta_D(\alpha=\frac{1}{4}, M=\frac{1}{4}) = 0.82
\label{a13}
\end{equation}  
These values almost coincide with those of the longitudinal structure
factor given in (\ref{a12}). \\
It is the purpose of this paper to determine the complete $\alpha$
dependence of the critical exponents $\eta_1(\alpha,M),
\eta_3(\alpha,M)$. \\
The paper is organized as follows: \\
In section 2 we exploit the range of validity of Marshall's sign
rule (\ref{a8}). In section 3 we study the impact of the next to
nearest neighbour coupling on the lowlying excitations and on the
static structure factors. The latter are computed numerically on
systems up to $N=32$. A finite size analysis of (\ref{a9}) and
(\ref{a11}) yields the critical exponents $\eta_1(\alpha,M),
\eta_3(\alpha,M)$. Section 4 is devoted to the study of an unexpected
phenomenon, which we found for negative $\alpha$-values: For $\alpha <
\alpha_-(M)$ the finite size behaviour of the longitudinal
structure factor (\ref{a11}) changes systematically from a monotonic
increase to a decrease.  
%In sections 2 and 3 we study the impact of frustration on the
%lowlying excitations and on the static structure factors. The latter
%are computed numerically on systems with $N=4,6,...,28$ sites. A
%finite size analysis yields the critical exponents $\eta_1(\alpha,M),
%\eta_3(\alpha,M)$. The singularities in the static structure factors
%are generated by infrared singularities in the dynamical structure
%factors. In other words, the system is critical at the momenta
%$p=\pi$, $p=p_1(M)$ and $p=p_3(M)$ and one might wonder whether its
%properties can be described by conformal field theory \cite{15}. Indeed, conformal
%field theory allows to determine the critical exponents $\eta_1(\alpha,M),
%\eta_3(\alpha,M)$ from the finite size dependence of the gap and the
%spinwave velocity \cite{16} \cite{17}. 
%This second determination is performed in section
%4.\\
%
%
%
%
%
\section{The impact of frustration on the ground states. Levelcrossings.}
It was pointed out in the introduction that the momenta 
$p_S(\alpha)$ of the ground state $|S,S_z=S,p_S(\alpha)>$ in the
sectors with total spin $S$ follow Marshall's sign rule for $\alpha \le
\frac{1}{4}$.
We found deviations from this rule for 
\begin{equation}
 \alpha > \alpha_0(M=\frac{S}{N})
\label{b1}
\end{equation}
We computed the ground state energies $E(\alpha,p_S(\alpha),M=\frac{S}{N},N)$ on
small systems 
with $N=10,..,20$ sites. The dependence on the frustration parameter
$\alpha$ is
shown in Fig. 1 for $N=12$. Here we have marked the different
ground state momenta by different symbols. \\
%The following features are worthwhile to note: 
At $M=\frac{S}{N}=0$ deviations from Marshall's sign rule occur
first at:
\begin{equation}
\alpha_0(M=0) = \frac{1}{2}
\label{b2}
\end{equation}
Here we meet the Majumdar Gosh \cite{9} model, which is known to have two
degenerate ground states, namely dimer states with momenta $p=0$ and
$p=\pi$, respectively. \\
A twofold degeneracy  - with respect to the ground state momenta
$p_S^{(1)} (\alpha), p_S^{(2)} (\alpha)$ emerges along the whole curve
$\alpha=\alpha_0(M)$, which is plotted for N=10,12,14,16,18,20 in
Fig.2 \\
The first momentum $p_S^{(1)} (\alpha_0)$ follows Marshall's sign rule
(\ref{a8}). 
We have looked for an empirical rule for the second momentum
$p_S^{(2)}(\alpha_0)$ but we did not find such a rule which holds for
all momenta and system sizes $N$. \\
%The 
%difference between the first and second momentum obeys an empirical
%rule:
%\begin{equation}
% \Delta p_S(\alpha_0) = |p_S^{(1)} - p_S^{(2)}| = \pi (1-2M)
%\label{b3}
%\end{equation}
%which is satisfied for $N=8,10,12$ with one exception, namely
%\begin{equation}
%p_S=\frac{2}{3} \pi \quad  S=\frac{N}{4} \quad \mbox{for} \quad N=12 \quad
%\mbox{and} \quad p_S=\frac{1}{5} \pi \quad S=\frac{N-2}{4} \quad \mbox{for} \quad N=10
%\label{b4}
%\end{equation}
In the saturating field limit $M \to \frac{1}{2}$ the curve
$\alpha_0(M)$ meets the point
\begin{equation}
\alpha_0(M \to \frac{1}{2}) = \frac{1}{4}
\label{b5}
\end{equation}
Indeed, the eigenvalue problem can be solved analytically for
$S=\frac{N}{2}-1$ with the ansatz (1 magnon states)
\begin{equation} 
|p, S=\frac{N}{2}-1> = \frac{1}{\sqrt{N}} \sum_x e^{ipx} |x>
\label{b6}
\end{equation}
where $|x>$ denotes a spin state with spin $-\frac{1}{2}$ at site $x$
and spin $\frac{1}{2}$  at all other sites.
The energy of the 1 magnon state is found to be:
\begin{equation}
E(\alpha,p,M=\frac{1}{2}-\frac{1}{N},N) =  2\cos p + \alpha 2 \cos
(2p) + (\frac{N}{2}-2) (1+\alpha)
\label{b7}
\end{equation}
The ground state energy and its momentum $p=p_S(\alpha)$ follows by
minimizing (\ref{b7}) with respect to $p$. For $\alpha < \frac{1}{4}$ the
ground state momentum is found to be $p_S^{(1)}(\alpha)=\pi, S=\frac{N}{2}-1$ in
accord with Marshall's sign rule. For $\alpha > \frac{1}{4}$ however, the
minimum is found for $p=p_S^{(2)}(\alpha)$ where
\begin{equation}
\cos p_S^{(2)}(\alpha) = \frac{1}{4\alpha} \hspace{5mm} S=\frac{N}{2}-1
\hspace{5mm} N\to \infty
\label{b8}
\end{equation}
On finite lattices, the difference between the two momenta turns out
to be:
\begin{equation}
\Delta p_S(\alpha_0=\frac{1}{4}) = |p_S^{(1)}(\alpha_0 = \frac{1}{4}) -
p_S^{(2)}(\alpha_0 = \frac{1}{4})| =\frac{2\pi}{N}  \hspace{5mm} S=\frac{N}{2}-1
\label{b9}
\end{equation} \\
As a consequence of the levelcrossing at
$\alpha=\alpha_0(M=\frac{S}{N})$, $M$ fixed, the derivatives of the
ground state energies 
\begin{equation}
\frac{\partial}{\partial \alpha} E(\alpha,p_S(\alpha),M=\frac{S}{N},N)
\label{b10}
\end{equation}
change discontinuously, as can be seen in an amplification of Fig. 1.
%%%%%
%
%
%
%
%
\section{Softmodes in the Excitation Spectrum and the associated
$\eta$-Exponents.} 
We have studied the finite-size dependence of the energy gaps
\begin{equation}
 \omega_{\Delta S}(\alpha,p,M,N) = 
E(\alpha, p=p_S+p, M=\frac{S+\Delta S}{N},N) -
E(\alpha,p=p_S,M=\frac{S}{N},N) 
\label{c1} \\
\end{equation}
for $\Delta S=0$ and $\Delta S=1$ in the domain $\alpha < \alpha_0(M)$ where
the ground state momentum follows Marshall's sign rule. In this regime
the gap $ \omega_{\Delta S=1} (\alpha, p=\pi,M,N) $
vanishes in the thermodynamical limit in such a way that the scaled
quantity 
\begin{equation} 
\lim_{N \to \infty} N \omega_{\Delta S=1} (\alpha, p=\pi,M,N) =
\Omega_1(\alpha,M) \quad \alpha < \alpha_0(M)
\label{c3}
\end{equation}
converges to a finite non-vanishing limit.
The same holds for the gap $\omega_{\Delta S=0} (\alpha,
p=p_3(M),M,N)$, $p_3(M)=\pi(1-2M)$, if the next to nearest neighbour
coupling $\alpha$ is positive:
\begin{equation} 
\lim_{N \to \infty} N \omega_{\Delta S=0} (\alpha, p=p_3(M),M,N) =
\Omega_3(\alpha,M)
\label{c4}
\end{equation}
For negative $\alpha$-values ($\alpha<\alpha_-(M)<0$) however, we
observe a tendency in the numerical data, which at least hints to the
emergence of a gap at the momentum $p=p_3(M)$:
\begin{equation} 
\lim_{N \to \infty}\omega_{\Delta S=0} (\alpha, p=p_3(M),M,N) =
\Delta_3(\alpha,M) \quad \alpha < \alpha_-(M) < 0
\label{c4a}
\end{equation}
as can be seen from Fig. 3.
It is hard to decide from the finite system results ($N=16,20,24$)
the exact position $\alpha=\alpha_-(M)$ where the gap (\ref{c4a})
opens. \\
In the gapless regimes, where (\ref{c3}) and (\ref{c4}) is valid, we
expect that the critical behaviour of the system is properly described
by conformal field theory.
This means in
particular that the ratios \hspace{1mm} [\onlinecite{13}-\onlinecite{17}]
\begin{equation}
2 \theta_a(\alpha,M) = \frac{\Omega_a(\alpha,M)}{\pi v(\alpha,M)},
\quad a=3,1 ,
\label{c5} 
\end{equation}
can be identified with the critical
exponents $\eta_a(\alpha,M)$:
\begin{equation}
2 \theta_a(\alpha,M) = \eta_a(\alpha,M), \quad a=1,3.
\label{c6}
\end{equation}
Here
\begin{equation}
v(\alpha,M) = \frac{1}{2\pi} \lim_{N\to\infty} N
(E(\alpha,p=p_S+\frac{2\pi}{N},M=\frac{S}{N},N) - E(\alpha,p_S,M=\frac{S}{N},N))
\label{c7}
\end{equation}
is the spinwave velocity. 
%The latter appear in the large N dependence (\ref{a9}) and (\ref{a11}) of
%the transverse and longitudinal structure factor at the momenta
%$p=\pi$ and $p=p_3(M)$, respectively.
For fixed values of $M$ $(M=\frac{1}{6},\frac{1}{4},\frac{1}{3})$ we
have determined the $\alpha$-dependence of $2\theta_a(\alpha,M), \quad
a=1,3 ,$ 
from the energy differences (\ref{c1}) (\ref{c7}) as they enter in the
ratios (\ref{c5}). The result can be seen from the solid curves in Figs
4a,b,c. The solid dots represent the determination of the critical
exponents $\eta_1(\alpha,M), \eta_3(\alpha,M)$ as they follow from a
fit of the form
(\ref{a9}), (\ref{a11}) to the finite system results ($N \le 32$).
Comparing the two determinations we come to the following conclusions:
\begin{enumerate}
\item The identity $2\theta_1(\alpha,M)=\eta_1(\alpha,M)$ for the critical
exponent in the transverse structure factor is well established for
$-0.5 < \alpha < 0.25$. \\
The same holds for the 
identity $2 \theta_3(\alpha,M) = \eta_3(\alpha,M)$ for the
critical exponent in the longitudinal structure factor in the interval 
$ \alpha_-(M) < \alpha < 0.25$.
If we approach the curve $\alpha=\alpha_0(M)$ the convergence of the
Lanczos algorithm slows down more and more, due to the emergence of
the level-crossing discussed in section 2.
\item The two curves $2\theta_1(\alpha,M)$, $2\theta_3(\alpha,M)$ cross each other
at $\alpha=\alpha_c(M)$
\begin{equation}
2\theta_1(\alpha_c(M),M) = 2\theta_3(\alpha_c(M),M) = 2\theta(M)
\label{c12}
\end{equation}
where 
\begin{equation}
\alpha_c(M=\frac{1}{6})=0.18 \quad \alpha_c(M=\frac{1}{4})=0.20 \quad
\alpha_c(M=\frac{1}{3})=0.24
\label{c13}
\end{equation}
and
\begin{equation}
2\theta(M=\frac{1}{6})=1.01 \quad
2\theta(M=\frac{1}{4})=1.02 \quad
2\theta(M=\frac{1}{3})=1.02
\label{c14}
\end{equation}
The $\alpha$-values are quite close to the transition point
$\alpha_c(M=0)=0.241$ from the spinfluid to the dimer phase. The same
holds for the critical exponents $\eta(M)$, which deviate only
slightly from $\eta(M=0)=1$.
\item The relation
\begin{equation}
4 \theta_1(\alpha,M) \theta_3(\alpha,M) =1
\label{c15}
\end{equation}
appears to be satisfied within a few percent for $0 \le \alpha \le \frac{1}{4}$.
\item For negative values of $\alpha$, we observe in the data for
$\eta_3(\alpha,M)$ a discontinuous structure (open symbols).
Looking at the numerical data, which enter in the determination of
$\eta_3(\alpha,M)$ via eq. (\ref{a11}), we found a systematic change
in the finite size dependence.
For $\alpha > \alpha_-(M)$
\begin{equation} 
\alpha_-(M=\frac{1}{6})= -0.31 \quad \alpha_-(M=\frac{1}{4})= -0.19
\quad \alpha_-(M=\frac{1}{3})= -0.15 
\label{c10}
\end{equation}
the longitudinal structure factor monotonically increases with $N$,
whereas it decreases for $\alpha < \alpha_-(M)$. 
In the latter regime we expect the emergence of the gap (\ref{c4a}).
\end{enumerate}
\section{The disappearance of a field dependent softmode}
The change in the finite size dependence of the gap (\ref{c4},
\ref{c4a}) and of $S_3(\alpha,p_3(M),M,N)$ provides us with a first
hint, that the field dependent softmode at $p=p_3(M)=\pi(1-2M)$ might
disappear for $\alpha<\alpha_-(M)$. In this section, we are looking
for further evidence for this hypothesis. In Figs 5a,b we compare the
momentum distribution of $S_3(\alpha,p,M=\frac{1}{4},N)$ for
$\alpha=\alpha_-(M=\frac{1}{4})=-0.19$ and $\alpha=-0.4$, respectively.
At $\alpha = \alpha_-(M=\frac{1}{4})= -0.19$ (Fig 5a) the momentum
distribution is well approximated by two straight lines with
different slopes for $p<p_3(M)$ and $p>p_3(M)$, respectively. This
discontinuity is more and more washed out, if the next to nearest
neighbour coupling decreases further. E.g. at $\alpha = -0.4$ (Fig 5b)
the $p$-distribution of the longitudinal structure factor appears to be
smooth in the thermodynamic limit. The approach to this limit is
indicated by an arrow. $S_3(\alpha,p,M,N), \alpha \le \alpha_-(M)$ is
monotonically decreasing with $N$ for $p \le p_3(M)$ but increasing for
$p>p_3(M)$. \\
A more drastic effect can be seen in the dynamical structure factor:
\begin{equation}
S_3(\alpha,\omega,p,M,N) = \sum_n \delta(\omega-(E_n-E_s))
|<n|S_3(p)|s>|^2
\label{d1}
\end{equation}
which we computed by means of the recursion method \cite{19} \quad \cite{20} 
for $M=\frac{1}{4}$ and $N=28$.
The excitation spectrum is plotted in Figs 6 a,b for $\alpha = -0.19$
and $\alpha = -0.4$, respectively. The numbers denote the
corresponding relative spectral weight in percentage terms. 
The curves guide the eye to
the excitations with dominant spectral weight.
For $\alpha = -0.19$ (Fig 5 a) the spectral weight is distributed over
a band of excitation energies which broadens in the vicinity of the
momentum $p=p_3(M)$. For $\alpha = -0.4$ (Fig 5b), however,
the spectral weight is more concentrated at higher excitation
energies. In particular, the lowest excitation at $p=p_3(M)$ has a
relative spectral weight less than $10 \%$ for $N=28$. 
%A finite size
%analysis of the corresponding transition probability yields:
%
%
%
%
\section{Discussion and Conclusion}
In this paper, we studied the impact of a next to nearest neighbour
coupling $\alpha$ and an external field $B$ on the zero temperature
properties of the one dimensional spin $\frac{1}{2}$ antiferromagnetic
Heisenberg model. We found the following features:
\begin{enumerate} 
\item The momentum of the ground state follows Marshall's sign rule
(\ref{a8}) for $\alpha \le \alpha_0(M)$ where
$\alpha_0(0)=\frac{1}{2}$ and $\alpha_0(\frac{1}{2}) =
\frac{1}{4}$. The ground state is twofold degenerate with respect to
its momentum for $\alpha=\alpha_0(M)$
\item A study of the finite size dependence (\ref{a9}) and (\ref{a11})
yields the $\alpha$-dependence of the critical exponents
$\eta_1(\alpha,M), \eta_3(\alpha,M)$ associated with the softmode
singularities at $p=\pi$ and $p=p_3(M)$ in the transverse and
longitudinal structure factor, respectively. Good agreement is found
with the prediction (\ref{c5}), (\ref{c6}) of conformal field theory
for $\eta_1(\alpha,M) \quad -\frac{1}{2} < \alpha < \frac{1}{4}$ and
for $\eta_3(\alpha,M) \quad 0 < \alpha < \frac{1}{4}$ (Figs 4a-c).
For these $\alpha$-values the spectral weight -- entering into the
definition of the corresponding dynamical structure factors (\ref{d1})
-- is concentrated around the lower bound of the excitation
spectrum. This seems to be a crucial condition in order that the
critical behaviour is described correctly by conformal field theory.
In the thermodynamical limit the dynamical structure factors
$S_1(\alpha,\omega,p=\pi,M)$ and $S_3(\alpha,\omega,p=p_3(M),M)$
develop infrared singularities $\omega^{-(2-\eta_a(\alpha))} \quad
a=1,3$, which can clearly be seen in a finite size scaling
analysis. Such an analysis was performed in Ref. \onlinecite{13} for $\alpha=0$. 
\item Deviations from the relation (\ref{c7}) -- predicted by
conformal field theory -- appear in the longitudinal case $a=3$ (Figs
4a-c) for negative values of the next to nearest neighbour coupling
and increasing $M$-values. This is accompanied by the fact that the
spectral weight in (\ref{d1}) is distributed over a band of excitation
energies, which broadens with decreasing values of $\alpha$.
\item There are several indications that the field dependent soft mode
at $p=p_3(M)=\pi (1-2M)$ disappears for negative next to nearest
neighbour couplings $\alpha < \alpha_-(M) < 0$: A gap (\ref{c4a}) opens
and the longitudinal structure factor (\ref{a11}) changes its finite
size dependence from a monotonic increase to a
decrease. Moreover the cusp-like singularity in the momentum
dependence at $p=p_3(M)$ is washed out and the spectral weight is
shifted from low to higher excitation energies. \\
Therefore, we find a further confirmation of the hypothesis --
formulated in Ref.\onlinecite{21} -- namely that field dependent soft modes only
exist if the system is sufficiently frustrated. As was pointed out in
Ref.\onlinecite{21} this condition is not satisfied in the twodimensional spin
$\frac{1}{2}$ antiferromagnetic Heisenberg model with nearest neighbour
coupling.
\end{enumerate}

%%%%%%%%%%%%%%%%%%%%%%%%%%%%%%%%%%%%%%%%%%%%%%%%%%%%%%%%%%%%%%%%%%%%%%%%%%%%%%%%
%
%
\newpage
\centerline{\bf Figure Captions}
%
%
%%%%%%%%%%%%%%%%%%%%%%%%%%%%%%%%%%%%%%%%%%%%%%%%%%%%%%%%%%%%%%%%%%%%%%%%%%%%%%%%
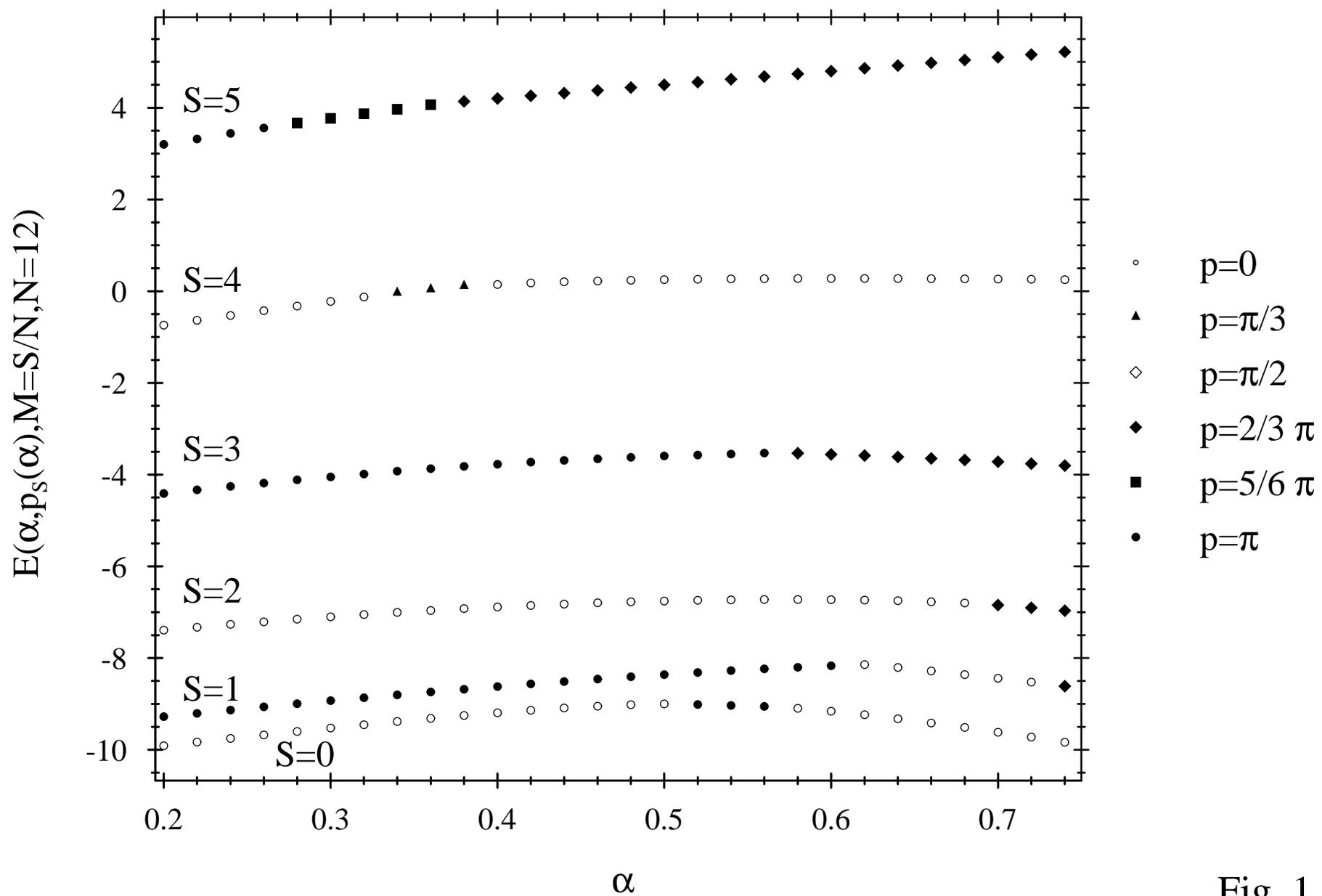
\begin{figure}
\caption[1]{
     The ground state energies $E(\alpha,p_S,M=\frac{S}{N},N)$ in the sector with
     total spin $S$ on a ring with $N=12$ sites. The ground state momenta
     $p_S(\alpha)$ change with the next to nearest neighbour coupling
     $\alpha$ as indicated by the different symbols.}
\end{figure}

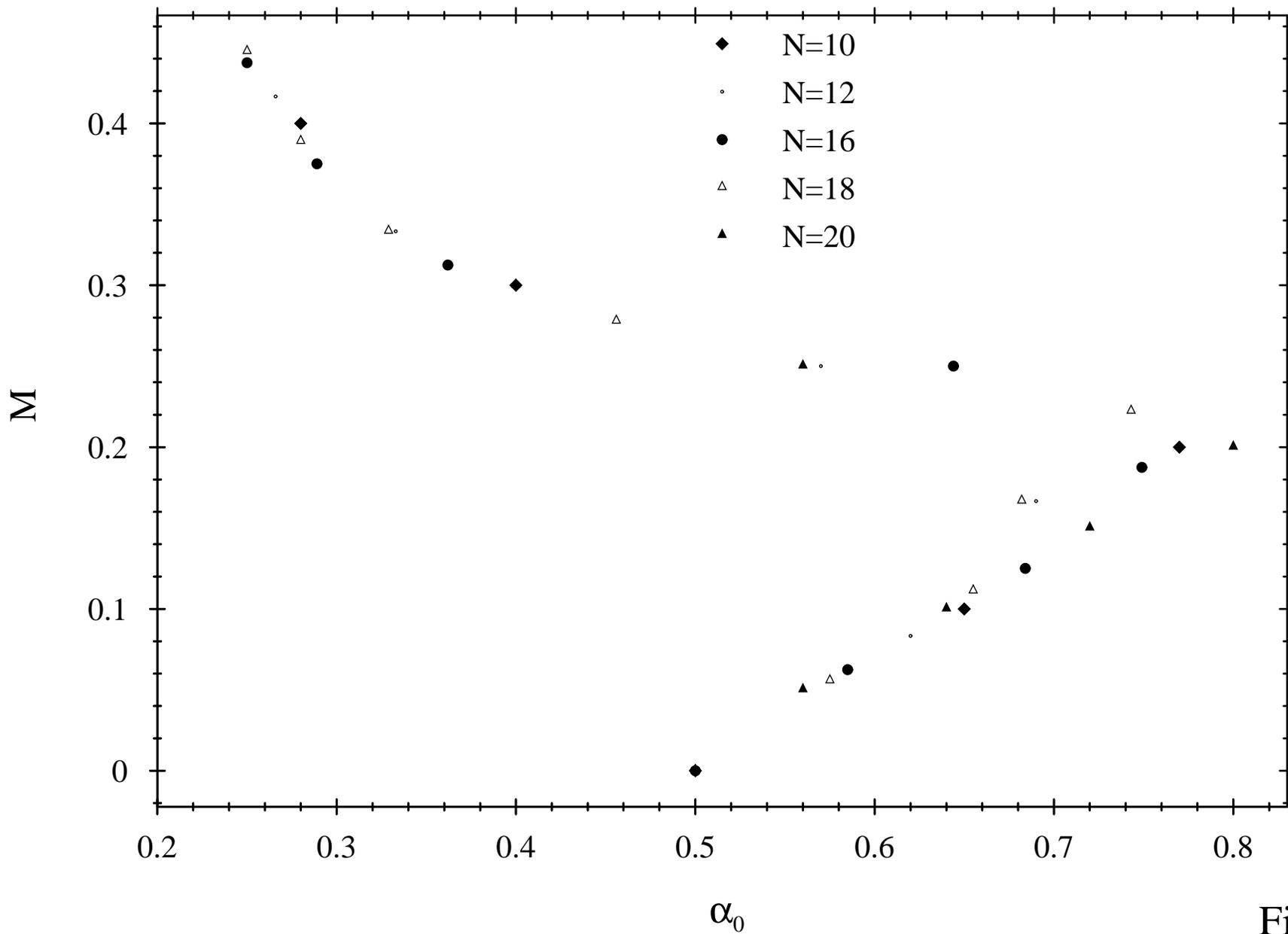
\begin{figure}
\caption[2]{ 
     The curve $\alpha=\alpha_0(M=\frac{S}{N})$ where the ground state
     $|S,p>$ in the sector with total spin $S$ is degenerate with respect
     to the momentum $p=p_S^{(1)}(\alpha), p_S^{(2)}(\alpha)$.}
\end{figure}

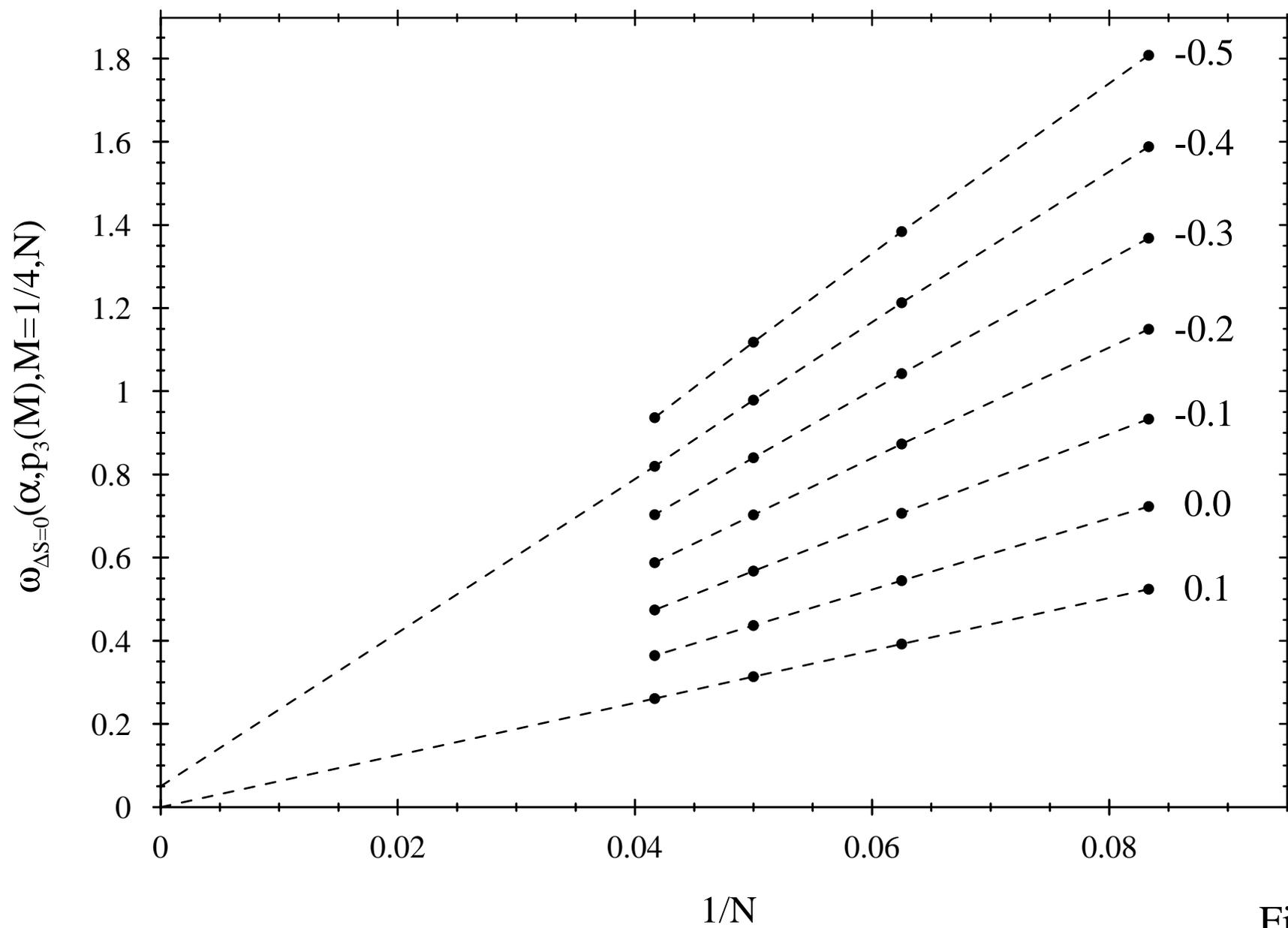
\begin{figure}
\caption[3]{ 
     Finite size dependence of the gap $\omega_{\Delta
     S=1}(\alpha,p_3(M),M,N)$ for $\alpha = 0.1, 0.0, -0.1, -0.2, -0.3,
     -0.4, -0.5$.}
\end{figure}

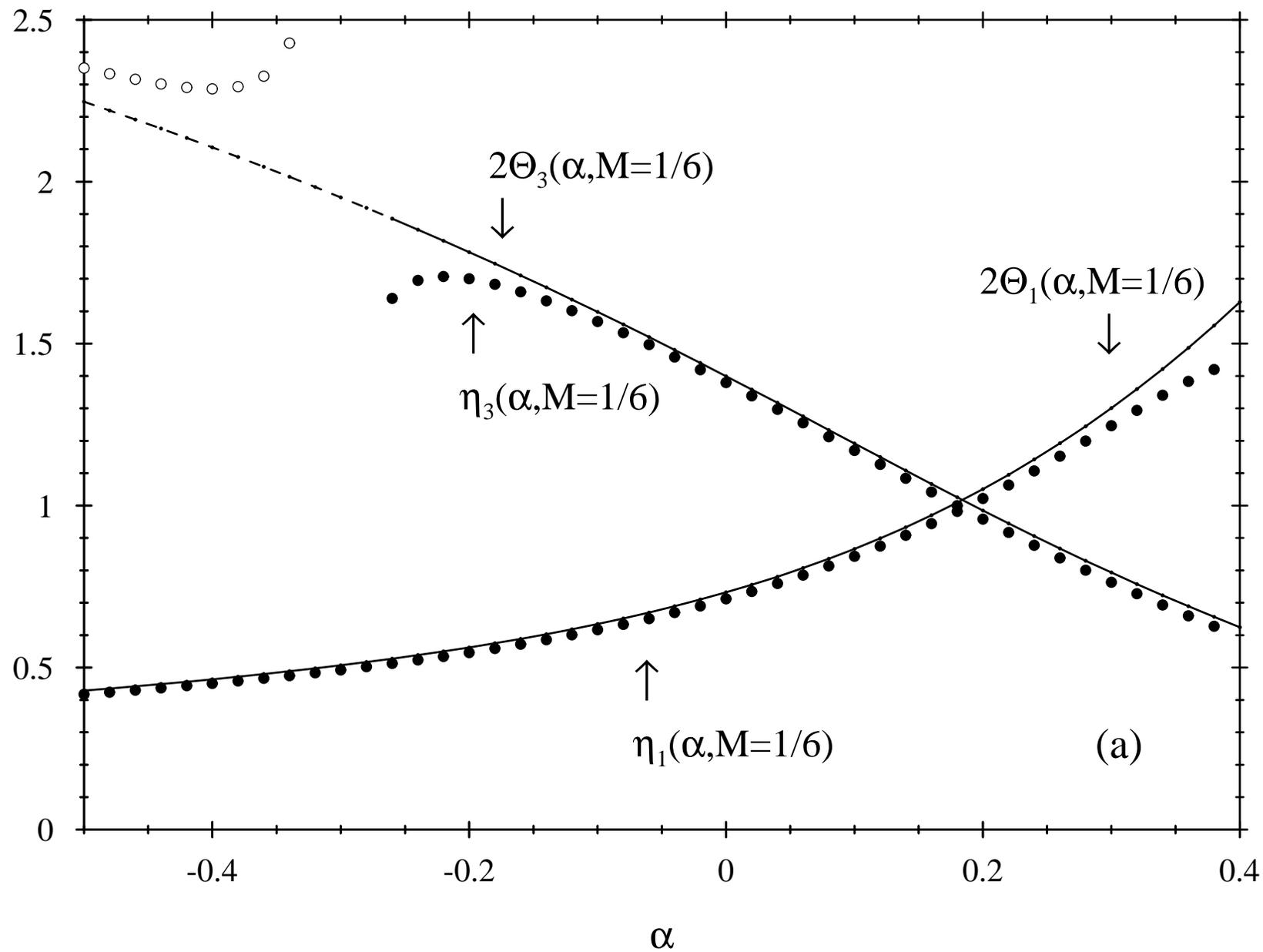
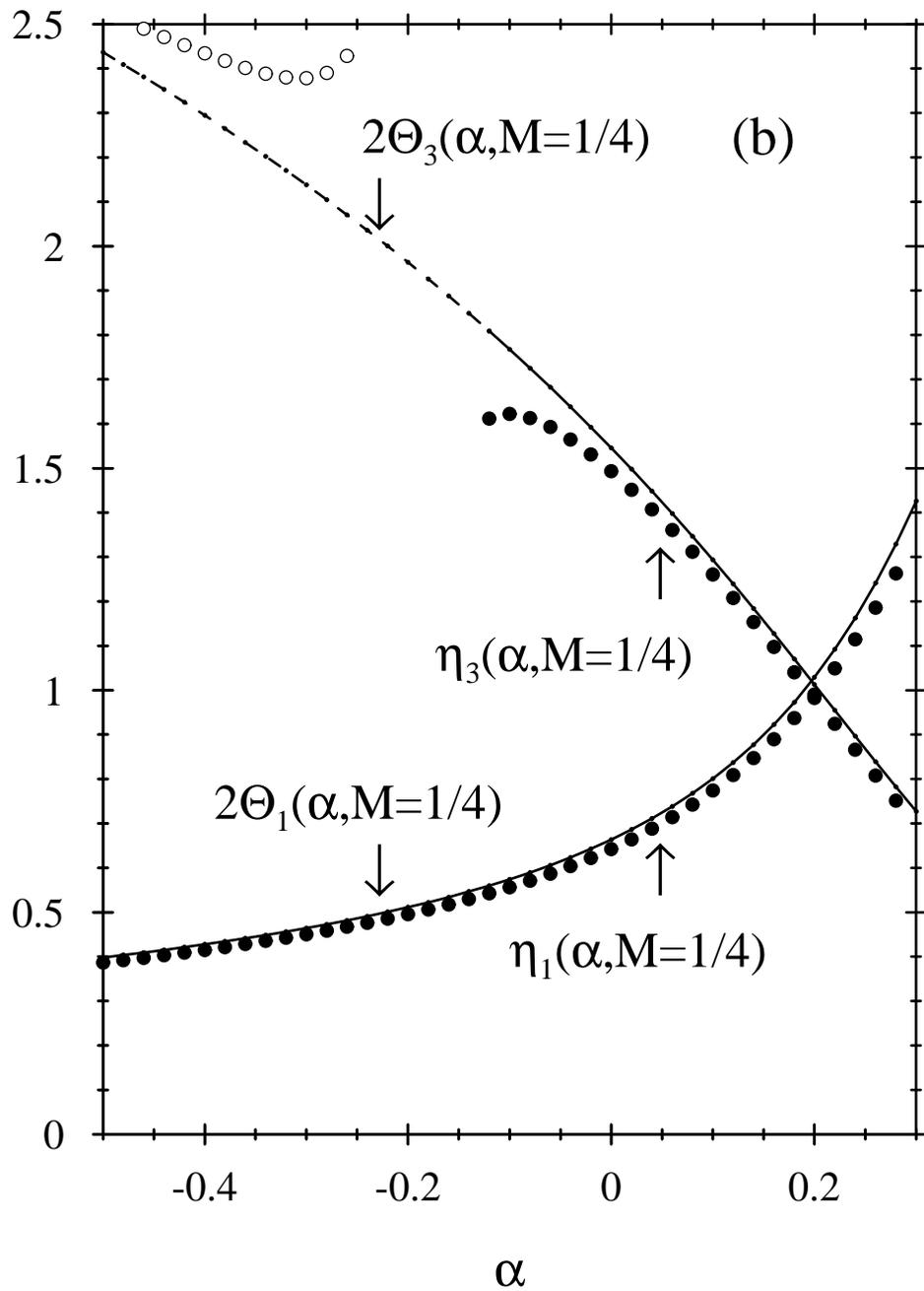
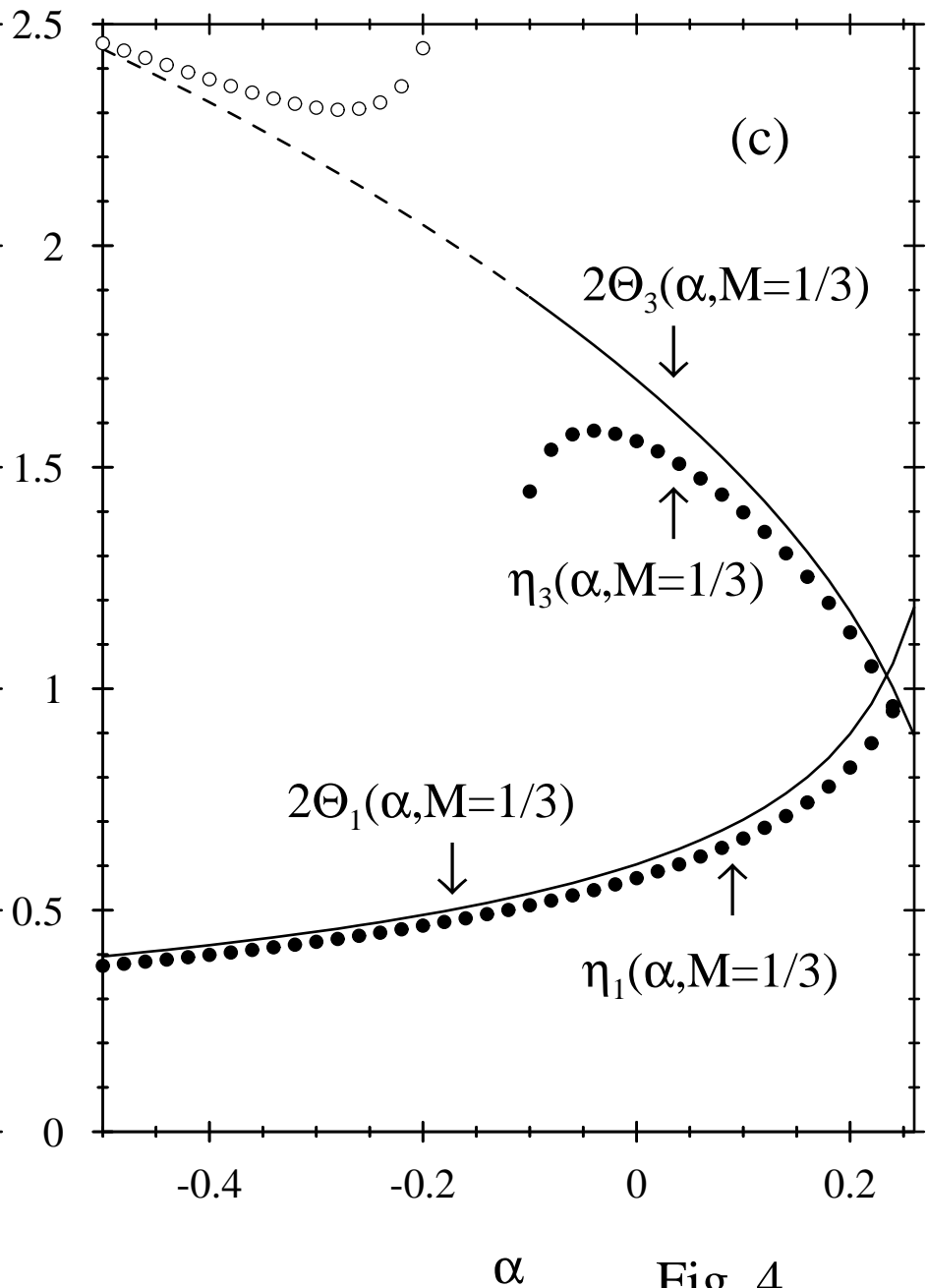
\begin{figure}
\caption[4]{ 
     Comparison of the ratio $2\theta_i(\alpha,M) \quad i=1,3$
     (\ref{c5}) \quad (solid
     curves) and the critical exponents $\eta_i(\alpha,M) \quad i=1,3$ in
     the static structure factors (\ref{a9})(\ref{a11}). 
     a) $M=\frac{1}{6}$  \quad  b) $M=\frac{1}{4}$ \quad c) $M=\frac{1}{3}$} 
\end{figure}

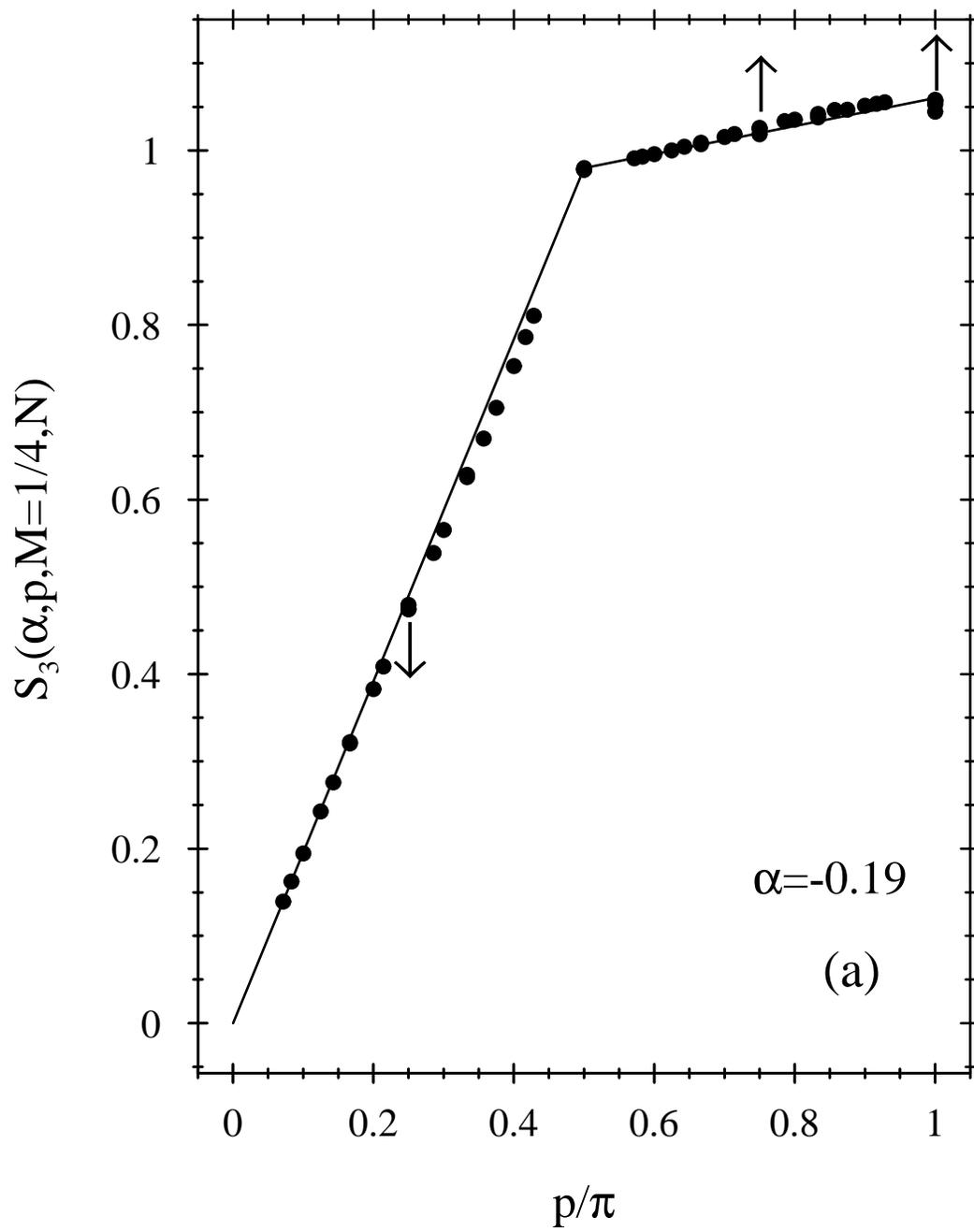
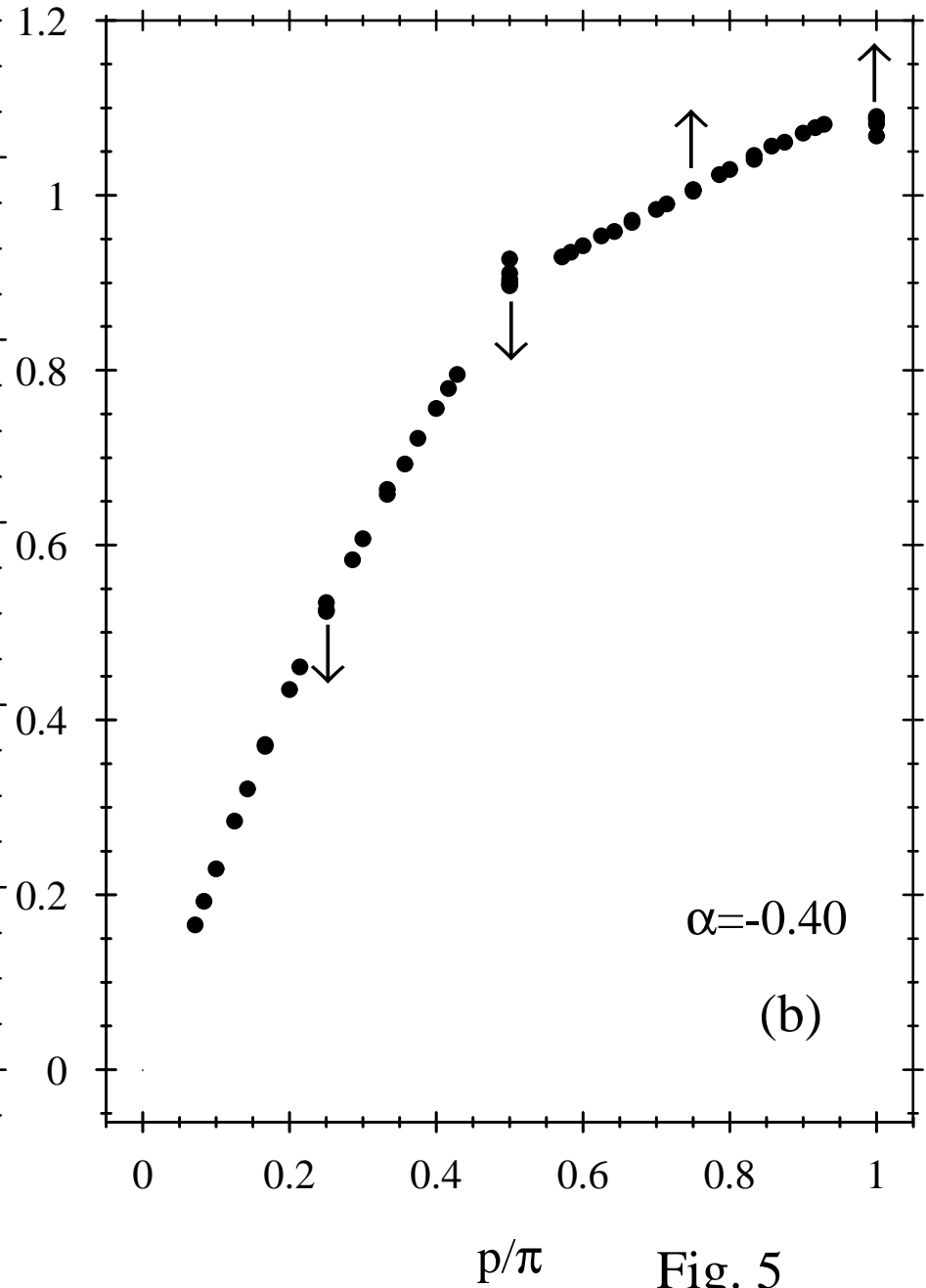
\begin{figure}
\caption[5]{ 
     The momentum dependence of the longitudinal structure factor
     $S_3(\alpha, p, M=\frac{1}{4},N) \quad N=28,24,20,...$ .
     a) $\alpha = \alpha_-(M=\frac{1}{4}) = -0.19$ \quad
     b) $\alpha = -0.40$}
\end{figure}

\begin{figure}
\caption[6]{ 
     Excitation energies and relative spectral weights in the
     dynamical structure factor
     $S_3(\alpha,\omega,p,M=\frac{1}{4},N=28)$ in percentage terms. 
     The lines connect the
     excitations with the dominant spectral weight. 
     a) $\alpha=\alpha_-(M=\frac{1}{4}) = -0.19$ \quad b) $\alpha = -0.4$}
\end{figure}

\section*{References}
\begin{thebibliography}{99}
%
\bibitem{1}
Haldane F D M 1982 {\sl Phys. Rev.} B {\bf 25} 4925; {\sl Phys. Rev.}
B {\bf 26} 5257 (erratum)
%
\bibitem{2}
Tonegawa T and Harada I 1987 {\sl J. Phys. Soc. Japan} {\bf 56} 2153;
1988 {\sl Proc. Int. Conf. on Magnetism, J.Physique coll. Suppl.} {\bf
49} C8 1411
%
\bibitem{3}
Igarashi J and Tonegawa T 1989 {\sl Phys. Rev.} B {\bf 40} 756; 1989
{\sl J. Phys. Soc. Japan} {\bf 58} 2147
%
\bibitem{4}
Kuboki K and Fukuyama H 1987 {\sl J. Phys. Soc. Japan} {\bf 56} 3126
%
\bibitem{5}
Affleck I, Gepner D, Schulz H J and Ziman T 1989 {\sl J. Phys. A:
Math. Gen.} {\bf 22} 511
%
\bibitem{5a} 
Tonegawa T and Harada I 1987 {\sl J. Phys. Soc. Japan} {\bf 56} 2153;
1988 {\sl Proc. Int. Conf. on Magnetism, J. Physique Coll. Suppl.}
{\bf 49} C8 1411
%
\bibitem{6}
Tonegawa T, Harada I and Igarashi J 1990 {\sl
Prog. Theor. Phys. Suppl.} {\bf 101} 513
%
\bibitem{7}
Tonegawa T, Harada I and Kaburagi M 1992 {\sl J. Phys. Soc. Japan}
{\bf 61} 4665
%
\bibitem{8}
Okamoto K and Nomura K 1992 {\sl Phys. Lett.} {\bf 169A} 433
%
\bibitem{9}
Majumdar C K and Ghosh D K 1969 {\sl J. Math Phys.} {\bf 10} 1388 \\
Majumdar C K 1970 {\sl J Phys. C: Solid State Phys.} {\bf 3} 911
%
\bibitem{10}
Schmidt M, Gerhardt C, M\"utter K-H and Karbach M  1996
{\sl J. Phys. Condens. Matter} {\bf 8} 553
%
\bibitem{11}
Yang C N and Yang C P 1966 {\sl Phys. Rev.} {\bf 150} 321 327
%
\bibitem{12}
Marshall 1955 {\sl Proc Roy. Soc.} (London) A 232, 48 
%
\bibitem{13}
Fledderjohann A, Gerhardt C, M\"utter K-H, Schmitt A and Karbach M 1996
{\sl Phys. Rev.} B  {\bf 54} 7168 
%
\bibitem{14}
Schmidt M 1996  PhD thesis University of Wuppertal (unpublished)
%
\bibitem{15}
Cardy J L 1986 {\sl Nucl. Phys.} B {\bf 270}, 186 
%
\bibitem{16}
Schultz H J and Ziman T 1986 {\sl Phys. Rev.} B {\bf 33} 6545
%
\bibitem{17}
Bogoliubov N M, Izergin A G and Korepin V E  1986 {\sl Nucl. Phys.} 
B 275 [FS 17] 687
%
\bibitem{18}
Bogoliubov N M, Izergin A G and Reshetikhin N Y 1987 {\sl J. Phys. A:
Math. Gen.} {\bf 20} 5361
%
\bibitem{19}
Fledderjohann A, Karbach M, M\"utter K-H and Wielath P 1995 {\sl
J. Phys.: Condens. Matter} {\bf 7} 8993 
%
\bibitem{20}
Viswanath V S, Zhang S, Stolze J and M\"uller G 1994 {\sl Phys. Rev.} B
{\bf 49} 9702; Viswanath V S and M\"uller G {\sl The Recursion Method
- Application to Many Body Dynamics} Springer Lecture Notes in Physics
Vol. 23 (Springer Verlag, New York, 1994)
%
\bibitem{21}
Yang M S and M\"utter K H 1996, to be published
\end{thebibliography}
\end{document}